\documentclass[english,12pt,times,doublespace]{article}

\usepackage{babel}
\usepackage[dvips]{graphicx}
\usepackage{ae}
\usepackage{amssymb,amsmath}
\usepackage{dsfont}
\usepackage{moreverb}
\usepackage[usenames,dvipsnames,svgnames,table]{xcolor}
\usepackage{bm}
\usepackage{natbib}
\usepackage{multirow}
\usepackage{tikz}
\usepackage{authblk} % allow to use \affil
\usepackage[labelfont=bf]{caption}
\usepackage[dvips,colorlinks,bookmarksopen,bookmarksnumbered,linkcolor=blue,citecolor=red,urlcolor=red]{hyperref}
\newcommand\BibTeX{{\rmfamily B\kern-.05em \textsc{i\kern-.025em b}\kern-.08em
T\kern-.1667em\lower.7ex\hbox{E}\kern-.125emX}}

\newdimen\R
\newdimen\S
\newcommand{\emptycircle}[1][1]{%
\begin{tikzpicture}[scale=#1]
    \draw (0,0) circle (0.5\S);
 \end{tikzpicture}\hspace{-3pt}
}
\newcommand{\emptytriangle}[1][1]{%
\begin{tikzpicture}[scale=#1]
    \draw (0,0) -- (\S,0) -- (0.5\S,\S) -- cycle;
  \end{tikzpicture}\hspace{-3pt}
}
\newcommand{\filltriangle}[1][1]{%
\begin{tikzpicture}[scale=#1]
    \draw[fill=black] (0,0) -- (\S,0) -- (0.5\S,\S) -- cycle;
  \end{tikzpicture}\hspace{-3pt}
}
\newcommand{\emptysquare}[1][1]{%
\begin{tikzpicture}[scale=#1]
    \draw (0,0) -- (\S,0) -- (\S,\S) -- (0,\S) -- cycle;
  \end{tikzpicture}\hspace{-3pt}
}
\newcommand{\emptypentagon}[1][1]{%
\begin{tikzpicture}[scale=#1]
  \draw (90:\R) 
    \foreach \x in {162,234,...,449} {
      -- (\x:\R)
    }-- cycle (0:\R);
\end{tikzpicture}}
\newcommand{\fillpentagon}[1][1]{%
\begin{tikzpicture}[scale=#1]
  \draw[fill=black] (90:\R) 
    \foreach \x in {162,234,...,449} {
      -- (\x:\R)
    }-- cycle (0:\R);
\end{tikzpicture}}
\newcommand{\emptydiamond}[1][1]{%
\begin{tikzpicture}[scale=#1]
  \draw (0,0) -- (0.5\S,0.5\S) -- (0,\S) -- (-0.5\S,0.5\S) -- cycle;
\end{tikzpicture}}

\begin{document}

%%%%%%%%%%%%%%%%%%%%%%%%%%%%%%%%%%%%%%%%%%%%%%%%%%%%%%%%%%%%%%%%%%%
%\runninghead{Martins R., Silva G.L. and Andreozzi V.}
%\title{Bayesian Joint Modeling of Longitudinal and Spatial Survival AIDS Data}
%\author{Rui Martins\affil{a}\corrauth, Giovani L. Silva\affil{b}\affil{c} and Valeska Andreozzi\affil{b}\affil{d}}
%\address{\affilnum{a}Centro de Investiga\c{c}\~{a}o Interdisciplinar Egas Moniz (CiCEM), Escola Superior de Sa\'{u}de Egas Moniz, Quinta da Granja, Monte de Caparica, 2829-511 Caparica, Portugal \\
%\corraddr{Escola Superior de Sa\'{u}de Egas Moniz, Quinta da Granja, Monte de Caparica, 2829-511 Caparica - Portugal. E-Mail: ruimartins@egasmoniz.edu.pt}
%%%%%%%%%%%%%%%%%%%%%%%%%%%%%%%%%%%%%%%%%%%%%%%%%%%%%%%%%%%%%%%%%%%

\title{Joint Dispersion Model with a Flexible Link}

\author{Rui Martins$^{1}$}

\footnotetext[1]{email:\texttt{ruimartins@egasmoniz.edu.pt} \\ Centro de Investiga\c{c}\~{a}o Interdisciplinar Egas Moniz (CiiEM), Escola Superior de Sa\'{u}de Egas Moniz, Quinta da Granja, Monte de Caparica, 2829-511 Caparica, Portugal.}

\date{April, 2016}
\maketitle %%%% for the non-format

%=============================================================
\begin{abstract}

The objective is to model longitudinal and survival data jointly taking into account the dependence between the two responses in a real HIV/AIDS dataset using a shared parameter approach inside a Bayesian framework. We propose a linear mixed effects dispersion model to adjust the CD4 longitudinal biomarker data with a between-individual heterogeneity in the mean and variance. In doing so we are relaxing the usual assumption of a common variance for the longitudinal residuals. A hazard regression model is considered in addition to model the time since HIV/AIDS diagnostic until failure, being the coefficients, accounting for the linking between the longitudinal and survival processes, time-varying. This flexibility is specified using Penalized Splines and allows the relationship to vary in time. Because heteroscedasticity may be related with the survival, the standard deviation is considered as a covariate in the hazard model, thus enabling to study the effect of the CD4 counts' stability on the survival. The proposed framework outperforms the most used joint models, highlighting the importance in correctly taking account the  individual heterogeneity for the measurement errors variance and the evolution of the disease over time in bringing new insights to better understand this biomarker-survival relation.

\end{abstract}

%--------------------- (Keywords) ----------------------

\maketitle

\newpage
%===============================================================
\section{Introduction}
\label{sec:introduction}

A joint model is, in simple terms, an approach to the construction and analysis of the joint distribution of a set of response variables, which may be of different types. In this work we are interested in simultaneously analyse longitudinal and survival data, taking advantage of the information that flows between these two data sources collected in the same patients. The incorporation of survival information into the longitudinal process it is somewhat equivalent to take into account the effect of an informative missing-data mechanism in order to assess the trends of the repeated measures. On the other hand, include adjusted longitudinal characteristics into the survival model improves the fit of the survival regression. It is this latter feature of joint models that will permit this work to reach its objectives. We intend to perform a survival analysis of the time-to-event since the diagnostic of HIV/AIDS infected individuals using simultaneously the longitudinal information coming from the CD4$^+$T lymphocyte counts (CD4 counts for short) among other covariates (\textit{see} section \ref{sec:data} for a description of the dataset).

In the context of a joint analysis the dynamics of the longitudinal repeated measures are usually postulated to belong in the class of the linear mixed-effects models with Gaussian errors \citep{Faucett1996,Wulfsohn1997,Rizopoulos2012}. Generally it is assumed that the residual variance for the individual longitudinal trajectory is common to all subjects. Few papers on joint modelling have been published considering a model for this source of variation, which means assuming different individual residual variances (dispersion modelling). \citet{Gao2011} is an example, where the within-subject variability is assumed to have a log-normal prior distribution. \citet{McLain2012} presents a dispersion model in a frequentist framework but we can see it as an hierarchical model, easily implemented in Bayesian terms, being the variation patterns modelled as a particular function of other variables (\textit{vide} section \ref{sec:model}). In our study, we resorted to this strategy with promising results (section \ref{sec:results}). Others authors, namely \citet{Jiang2015} assumed a latent class model for that variability and \citet{Chen2015} and \citet{Huang2014} modelled the error process with a skew-$t$ distribution.

One of the investigators' biggest concerns in this field is how and what characteristics should be shared by the longitudinal and survival processes. \citet{Rizopoulos2011} discuss this problem by means of general families of parameterisations describing the main features of the relationship between the two processes. The most used approach is to share a set of individual random effects believed to be the basis of a latent relationship. Generally those effects are used to explain the differences between the population mean longitudinal response and the individual-specific ones in a mixed model representation. We will extend this vision proposing the individual-specific longitudinal standard deviation as a covariate for the hazard model.

The most popular choices for the baseline hazard function specification are the parametric forms (\textit{e.g.} Weibull - \citet{Guo2004}) or the piecewise constant approximations \citep{Tang2014a,Baghfalaki2015}. In seeking the most flexibility as possible we will use an approach rooted on penalized cubic B-spline functions (P-Splines for short).

The majority of the joint analysis assume that the relationship between the two processes has the same strength over time, i.e. the coefficients of the survival model accounting for the effect of a particular longitudinal characteristic are time-invariant. But if we believe, for example, that an initial significant relationship will become non-significant some time later (or \textit{vice-versa}), what we are assuming is that the effect of the shared longitudinal characteristics on survival varies with time. We address this aspect using time-dependent coefficients approximated by P-Splines. Despite having a moderately use in survival analysis (\citet{Hennerfeind2006,Hofner2011}) time-varying coefficients are not a first option in joint models. Fortunately there are some exceptions. For instance \citet{Yu2014} use Splines to model these coefficients in a frequentist context; \citet{Hanson2011} rely on a linear combination of Gaussian kernels; \citet{Song2008b} use first order Taylor expansions of the function representing the time-varying coefficients.

The rest of the paper is organized as follows: section \ref{sec:model} describes the theoretical framework to fit a linear mixed effects model incorporating patient-specific variance, as well the survival model with time-dependent coefficients for the time to death and the linking structure; section \ref{sec:estimation} describes the estimation of this time dependent coefficients and the baseline hazard; the next (section \ref{sec:data}) highlights the data analysis and the application of the proposed model, being the results presented and assessed. Finally we give a discussion of the paper (section \ref{sec:discussion}) and supplementary material with the BUGS code to implement the models (section \ref{sec:suppmat}).

%=============================================================
\section{Model specification}
\label{sec:model}

Let us consider a study where an individual should come to an health center (hospital, clinic, laboratory, etc.) periodically in order to perform some blood tests (longitudinal process) and we follow him until he experiences the event of interest (failure) or be censored (lost to follow-up or end of the study). The longitudinal process for the $i$th individual, $m_i(t)$, $i=1,\ldots,N$, is measured with error. So, the observed one is $y_i(t)=m_i(t) + e_i(t)$, $t >0$. 

The $i$th vector of the $n_i$ observed repeated measures is defined as $\mathbf{y}_{i}\!=\!(y_{i}(t_{i1}),\ldots,\allowbreak y_{i}(t_{in_i}))\! \equiv \! (y_{i1},\ldots,y_{in_i})$ being $\mathbf{t}_{i}\! = \!(t_{i1},\ldots,t_{in_i})$ a vector of fixed individual times. This encompasses the possibility of having $N$ different measurement schedules and follow-up times. We take $T_{i}$ to be the observed (possibly right censored) time-to-event, for the $i$th individual and $\delta_i$ to be the failure indicator. The observed data without covariates for the $N$ independent subjects is $\mathcal{D}=\{\mathcal{D}_i\}_{i=1}^N = \{(\mathbf{y}_i,\mathbf{t}_i,T_i,\delta_i)\}_{i=1}^N$.

%---------------------------------------------------------------
\subsection{Longitudinal mixed dispersion model}
\label{sec:longitudinalmodel}
 
Consider that the longitudinal outcomes, $y_{ij}$'s, are described by a mixed effects dispersion model \citep{McLain2012},
\begin{align}
\label{eq:longitudinal}
y_{ij}|\bm{b}_i,\sigma^2_i \sim & \; \mathcal{N}\!\left(m_{i}(t_{ij}),\sigma^2_i\right)\!, \quad j=1,\ldots,n_i \\ 
m_{i}(t_{ij}) = & \; \bm{\beta}_1^\top \mathbf{x}_{1i}(t_{ij})   + \bm{b}_{1i}^\top  \mathbf{w}_{1i}(t_{ij}),  \\
 \sigma_i^2 = & \; \sigma_0^2 \exp\{\, \bm{\beta}_2^\top\mathbf{x}_{2i}(t_{ij}) + \bm{b}_{2i}^\top \mathbf{w}_{2i}(t_{ij}) \,\},
\label{eq:dispersionmodel}
\end{align}
where $\mathbf{x}_{1i}$, $\mathbf{x}_{2i}$, $\mathbf{w}_{1i}$ and $\mathbf{w}_{2i}$ are appropriate subject-specific vectors of covariates (possibly time-dependent); $\bm{\beta}_1$ and $\bm{\beta}_2$ are vectors of population regression parameters; $(\bm{b}_{1i}^\top,\bm{b}_{2i}^\top)=\bm{b}_i | \Sigma \sim \mathcal{N}_p(\bm{0},\Sigma)$ are time-independent subject-specific random effects capturing the inherent biological variability between individuals and allowing for between-individual heterogeneity in the mean and variance.

\begin{figure}[!htb]
\centering
\includegraphics[scale=0.5]{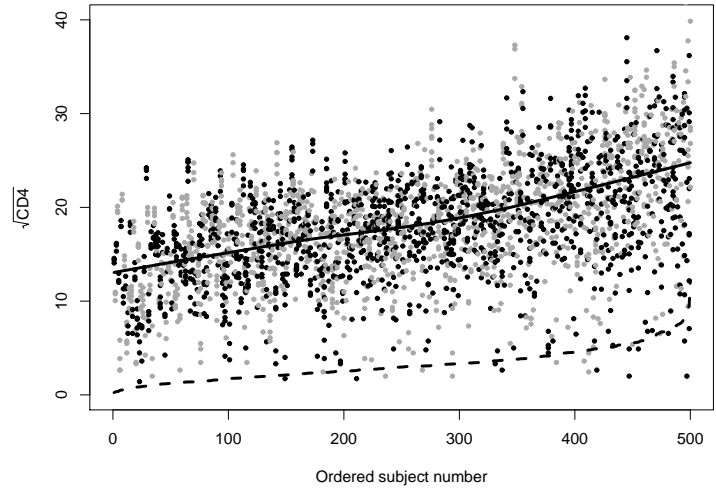}
\caption{Plot of the $\sqrt{\text{CD4}}$ values against the subject number resulting from ordering them according to the standard deviation of their $\sqrt{\text{CD4}}$. The $\sqrt{\text{CD4}}$ values for one subject are plotted using the same color and two adjacent subjects have different colors. The solid line represents the \textit{lowess} smooth of the $\sqrt{\text{CD4}}$ values and the dashed one is simply the connection between the 500 standard deviations.}
\label{fig:CD4vssubseqnumber}
\end{figure}

One of the common and major assumptions in longitudinal models is that the within-individual variances are homogeneous, although this is not always satisfied. For example, considering our dataset (section \ref{sec:dataset}), the plot of the individual $\sqrt{\text{CD4}}$ values against the standard deviation (Figure \ref{fig:CD4vssubseqnumber}) suggests considerable within-subject variance heterogeneity. Large values of the mean are associated with high variability. In equation \eqref{eq:dispersionmodel} the residual variance, $\sigma_i^2$, is assumed to be an individual property allowing for heterogeneity in the variance trends among the individuals. Modelling it and identify covariates related to this discrepancies seems wise. Particularly, in many HIV/AIDS studies, where investigators are interested in understanding the trends of the variability, having an individual estimate of the subject-residual variance can be a plus in the assessment of whether individuals with different biomarker's stability have different prognosis. %Lin 1997 p.1 and Lyles 1999 p.4

The representation in \eqref{eq:dispersionmodel}, where $\sigma_0^2$ acts as a ``baseline'' variance, allows for both increasing and decreasing variance across individuals. $\mathbf{x}_{2i}$ and $\mathbf{w}_{2i}$ represent the variables influencing the within-subjects variance. In this way, one can examine whether contextual variables are related to the within-individual variance. %Hedeker Mermelstein 07.pdf
In addition, this specification can be considered as an extension to the dispersion model defined in \citet{Lin1997} or the one defined by \citet{Gao2011}. The former modelled the individual-specific measure of stability, $\sigma_i^2$, through an inverse gamma prior distribution, which is a special case of \eqref{eq:dispersionmodel}. The latter do not consider possible covariate effects. Indeed, if we consider $\exp\{\, \bm{\beta}_2^\top\mathbf{x}_{2i}(t_{ij}) + \bm{b}_{2i}^\top \mathbf{w}_{2i}(t_{ij}) \,\}=1$ in \eqref{eq:dispersionmodel} we have $\sigma_i^2=\sigma_0^2$ and we are back to the simple models where $\sigma_i^2$ accounts for the randomness in stability by using some prior distribution for this parameter. More details about the choose of this prior distribution are discussed in section \ref{sec:priorspecifications}.

%---------------------------------------------------------------
\subsection{Hazard model with time-varying coefficients}
\label{sec:survivalmodel}

Various approaches have been proposed to link the longitudinal and survival processes. In this work, we tackle joint modelling considering the inclusion of some characteristics of the longitudinal trajectory, namely the random effects, $\bm{b}_i$, and the standard deviation, $\sigma_i$, into a hazard model (shared parameters approach), described below,
\begin{equation}
h_i(t \mid \bm{b}_i,\sigma_i) = h_0(t)\exp\{\; \bm{\beta}_3^\top \mathbf{x}_{3i} + \mathcal{C}_i\{\bm{b}_i,\sigma_i;\bm{g}(t)\} \; \} = h_0(t) \exp\{\varrho_i(t)\},
\label{eq:hazard}
\end{equation}
where $\mathbf{x}_{3i}$ is a subject-specific vector of baseline covariates and $\bm{\beta}_3$ is the respective population parameters vector. $\mathcal{C}_i\{.\}$ is a function specifying which components of the longitudinal process are directly
related to $h_i(.)$. Finally, $\bm{g}(t)\!=\!(g_1(t),\ldots,g_L(t))$ is an appropriate vector, set case-by-case, of $L$ smooth functions (approximated by P-Splines - see section \ref{sec:coefficientsestimation}) representing the time-varying regression coefficients \citep{Hennerfeind2006}, which measure the effect of some particular characteristics of the longitudinal outcome to the hazard. These coefficients are particularly useful in explaining the effect of a time-independent covariate on survival when its impact is not constant throughout the time, \textit{e.g.} if an initially significant relation becomes non-significant after a certain period. Baseline hazard, $h_0(t)$, can have a parametric form (\textit{e.g.} Weibull) or be specified using P-Splines or a piecewise constant function.

%=============================================================
\section{Estimation, specification and model choice}
\label{sec:estimation}

In this section we present the estimation of the time-varying coefficients, $g_l(t)$, $l=1,\ldots, L$, the different forms that the baseline hazard function, $h_0(.)$, can have and the specification of $\mathcal{C}_i\{.\}$, as well as the posterior distribution of the joint model. Finally, we will allude to the Watanabe-Akaike information criterion (WAIC, \cite{Watanabe2010}) for model choice.

%---------------------------------------------------------------
\subsection{Time-varying coefficients}
\label{sec:coefficientsestimation}

To fit model \eqref{eq:hazard} we will apply a penalized spline regression method to the elements of the vector $\bm{g}(t)$. The main idea of a regression via splines is described as follows and more details can be found in \cite{Lang2004}, \cite{Brezger2006} and \cite{Hennerfeind2006}. Using a linear combination of $Q_l$ B-spline basis functions, $\{B_{lq}(t)\}_{q=1}^{Q_l}$, $l=1,\ldots,L$, defined over a grid of $s_l+1$ equally spaced knots, $t_{\min}\!=\!\kappa_{l0}\!<\!\kappa_{l1}\!<\!\ldots\!<\!\kappa_{ls_l}\!=\!t_{\max}$, over the domain of $t$, we will be able to well approximate $g_l(t)$. As in \cite{Eilers2010} all the basis functions have the same shape including the boundaries of the domain, contrasting, for example, with natural B-splines, where near the boundaries the basis functions have different shapes. In our case the splines are of order $p=3$ and computed as differences of truncated power functions \citep{Eilers2010}. Thus, the expansion of $g_l(t)$ into B-spline basis is
\begin{equation}
 g_l(t)=\sum_{q=1}^{Q_l}{\gamma_{lq}B_{lq}(t)},\quad l=1,\ldots,L,
\label{eq:gsplines}
\end{equation}
where $Q_l=s_l+3$, $\bm{\gamma}_l=(\gamma_{l1},\ldots,\gamma_{lQ_l})$ are unknown vectors of fixed coefficients and $B_{lq}(t)$ denotes the $q$th basis function of the B-Spline. From know on we will drop index $l$ from the notation of the $Q_l$, $\kappa_{l}$, $s_l$, and $B_{lq}(t)$, because we are assuming  the same basis for all the B-Splines associated with each $g_l(t)$. To ensure that our functions $g_l(t)$ are smoothed enough, we penalize the first order differences between adjacent elements in $\bm{\gamma}_l$ using a first order random-walk to define their priors, i.e.,
\begin{equation}
\gamma_{l1} \sim \mathcal{N}(0,1000), \;\; \gamma_{lq}\mid\tau^2_l \sim \mathcal{N}(\gamma_{l,q-1},\tau^2_l),\quad l=1,\ldots,L;\; q=2,\ldots,Q,
\label{eq:priorrandomwalk}
\end{equation}
where $\tau^2_l$ acts as the variance controller. Large values correspond to a rougher curve. First order penalty aims to avoid abrupt jumps between adjacent elements in $\bm{\gamma}_l$. Instead a second order random-walk could be used to avoid departures from a linear trend as in \cite{Lang2004} and \cite{Kneib2007}.

Joint prior of the elements in $\bm{\gamma}_l$ is defined as a product of conditional densities defined by \eqref{eq:priorrandomwalk} yielding a Multivariate Gaussian distribution
\begin{equation}
\pi(\bm{\gamma}_l \mid \tau^2_l) \propto \tau^{-(Q-1)}_l\exp\left\{ -\frac{1}{2\tau^2_l} \bm{\gamma}^\top_l \mathbf{P} \bm{\gamma}_l\right\},\quad l=1,\ldots,L;\; q=1,\ldots,Q.
\label{eq:multivariatgaussbeta}
\end{equation}
$\mathbf{P}=\mathbf{D}^\top \mathbf{D}$ is the precision matrix and $\mathbf{D}$ is the $(Q-1)\times Q$ matrix of the first order differences.

%---------------------------------------------------------------
\subsection{Baseline hazard}
\label{sec:basehazestimation}

Various choices have been made to deal with $h_0(t)$. In the traditional Cox model \citep{Cox1972} it is treated as a nuisance parameter and left unspecified, relying the estimation process on the partial-likelihood function \citep{Costa2009}.
However, a completely unspecification can cause underestimation \citep{Hsieh2006}. On the other hand we have those who completely specify it using a parametric form (\textit{e.g.} Weibull) as in \cite{Guo2004}. Other popular choices are specifications by splines (\cite{Rizopoulos2009} and \cite{Kneib2007}) or a piecewise constant function (\cite{Tseng2005}, \cite{Zhu2012} and \cite{Rizopoulos2011} just to name a few). A baseline specification provides the possibility to direct estimate the full survival time distribution contrary to what happens when we left $h_0(t)$ unspecified.

Assuming that each $T_i$, $i=1,\ldots,N$, has a Weibull distribution, $\mathcal{W}(\rho,e^{\varrho_i(t)})$, $h_i(t|.)$ in \eqref{eq:hazard} is defined as
\begin{equation}
h_i(t \mid \bm{b}_i,\sigma_i) = \rho t^{\rho-1}\exp\{\varrho_i(t)\}.
\label{eq:hazardweibull}
\end{equation}
In case of a spline approach $h_i(t|.)$ is defined as
\begin{equation}
h_i(t \mid \bm{b}_i,\sigma_i) = \exp\{ g_0(t) + \varrho_i(t) \},
\label{eq:hazardspline}
\end{equation}
where now $\bm{g}(t)\!=\!(g_0(t),g_1(t),\ldots,g_L(t))$ accommodates  $g_0(t)=\log[h_0(t)]$, which is approximated in a similar manner to that of the $g_l(t)$ coefficients, $l=1,\ldots,L$, in section \ref{sec:coefficientsestimation}. Finally, if $h_0(t)$ is a piecewise constant function defined over a finite partition of the time, $0\!=\!a_1,a_2,\ldots,a_K,a_{K+1}=\infty$, we define the baseline hazard as $h_0(t)=\sum_{k=1}^K{\lambda_k} \mathds{1}_{[a_k,a_{k+1}[}(t)$, where $\bm{\lambda}=(\lambda_1,\ldots,\lambda_K)$ is a vector of positive but unknown parameters, each one representing the constant local hazard for $t \in [a_k,a_{k+1}[$, $k=1,\ldots,K$. Thus we can write 
\begin{equation}
h_i(t \mid \bm{b}_i,\sigma_i) = \sum_{k=1}^K{\lambda_k} \mathds{1}_{[a_k,a_{k+1}[}(t)\exp\{ \varrho_i(t) \}.
\label{eq:hazardpiecewise}
\end{equation}

%---------------------------------------------------------------
\subsection{The posterior distribution}
\label{sec:priorsandposterior}

To form the contribution of the $i$th individual to the likelihood we assume a non-informative right censoring, events and censoring are independent of the process measurement schedule, the longitudinal and survival processes are independent given the random effects, $\bm{b}_i$, and, in addition, the elements of the vector $\mathbf{y}_i$ are assumed independent given the same set of random effects. Thus, the posterior distribution, $\pi(\bm{\theta}|\mathcal{D})$, will be proportional to
\begin{align}
 & L_{1i}(\bm{\theta} \mid \mathcal{D}_i) \times L_{2i}(\bm{\theta} \mid \mathcal{D}_i) \times \pi({\bm{\theta}}) \nonumber \\
  = & \left\{ \prod_{j=1}^{n_i}{f(y_{ij} | \bm{b}_i,\sigma_i^2)} \right\} \times \left\{ \; h(T_i |\bm{b}_i)^{\delta_i}S(T_i|\bm{b}_i)\;\right\}\times \pi({\bm{\theta}}),
\label{eq:likelihood}
\end{align}
where $\pi(\bm{\theta})$ denotes the prior distribution of the full parameters vector including the random effects, $\bm{b}_i$; $L_{1i}(\bm{\theta} | \mathcal{D}_i)$ is the $i$th individual longitudinal contribution, being $f(.)$ the Gaussian probability density function specified as
\begin{equation}
 f(y_{ij}| \bm{b}_i,\sigma_i^2) = \frac{1}{\sqrt{2\pi\sigma_i^2}}\exp\left\{-\frac{\left[y_{ij} - m_i(t_{ij})\right]^2}{2\sigma_i^2}\right\},
\label{eq:fyij}
\end{equation}
and finally $h(.)$ and $S(.)$ are, respectively, the hazard and the survival function for $T_i$.

All of our joint models will have the same longitudinal specification \eqref{eq:longitudinal}. On the other hand, hazard models \eqref{eq:hazardweibull}, \eqref{eq:hazardspline} and \eqref{eq:hazardpiecewise} will define three different joint models, implying three different specifications for the $i$th individual survival contribution, $L_{2i}(\bm{\theta} | \mathcal{D}_i)$. Thus, considering \eqref{eq:hazardweibull}
\begin{equation}
L_{2i}(\bm{\theta} \mid \mathcal{D}_i) = \left\{\! \rho T_i^{\rho-1} e^{\varrho_i(T_i)} \! \right\}^{\delta_i} \exp\!\left\{ - e^{\varrho_i(T_i)}\, T_i^{\rho} \right\}\!,
\label{eq:hSweibull}
\end{equation}
assuming \eqref{eq:hazardspline} 
\begin{equation}
L_{2i}(\bm{\theta} \mid \mathcal{D}_i) = \left\{\! e^{g_0(T_i) + \varrho_i(T_i) \!} \right\}^{\delta_i} \exp\!\left\{\! - \! \int_0^{T_i} \!\! e^{g_0(u) + \varrho_i(u) } \, du\right\}\!,
\label{eq:hSsplines}
\end{equation}
and finally, postulating \eqref{eq:hazardpiecewise} we have
\begin{equation}
L_{2i}(\bm{\theta} \mid \mathcal{D}_i) = \left\{\! \lambda_{\mathring{k}} e^{\varrho_i(T_i)} \!\right\}^{\delta_i} \exp\!\left\{\! - \left[\lambda_{\mathring{k}}(T_i - a_{\mathring{k}}) + \sum_{k=1}^{\mathring{k}-1} \lambda_k(a_{k+1} - a_k) \right] \! e^{\varrho_i(T_i)} \!\right\}\!,
\label{eq:hSpiecewise}
\end{equation}
where $\mathring{k}$ is the largest integer such that $a_{\mathring{k}} \leq T_i$.

Although we are working on a Bayesian framework we still have to evaluate the integral in \eqref{eq:hSsplines}. In our models we used a 15-points Gauss-Kronrod quadrature rule as in \cite{Rizopoulos2011}, but other numeric methods could be applied. \cite{Ibrahim2004} show an alternative to this approximation tool.

%--------------------------------------------------------------
\subsection{Model choice}
\label{sec:modelchoice}

The purposes of model comparison and selection will be achieved with a recent measure. The Watanabe-Akaike information criterion (WAIC, \cite{Watanabe2010}), or the widely applicable information criterion in its own words, is a measure of predictive accuracy and invariant to reparameterisations contrary to the DIC \citep{Spiegelhalter2002}. In addition WAIC is fully Bayesian and closely approximates Bayesian cross-validation. DIC is considered not fully Bayesian, in part because it is based on a point estimate. Following \cite{Gelman2014} we define the WAIC value as
\begin{align}
\text{WAIC} = & -2\left[\text{lppd} - p_\text{WAIC}\right] \nonumber \\ 
 = & -2\left[\log	\left\{ \prod_{i=1}^N p(\mathcal{D}_i|\mathcal{D}) \right\}  - \sum_{i=1}^N \mathbb{V}_{\bm{\theta}|\mathcal{D}} \! \left[ \log \{p(\mathcal{D}_i|\bm{\theta})\} \right] \right]\!.
\label{eq:WAIC}
\end{align}
Like the DIC models with smaller WAIC values should be preferred. More details about this criterion, including the computation in the MCMC context, can be found in the SuppMaterial.

%=============================================================
\section{HIV/AIDS data analysis}
\label{sec:data}

%-------------------------------------------------------------
\subsection{Dataset}
\label{sec:dataset}

Brazilian National AIDS Program generated three major electronic databases \citep{Fonseca2010}: (i) SINAN-AIDS (Information System for Notifiable Diseases of AIDS Cases) which is the most important electronic AIDS surveillance database, with all cases reported since 1980; (ii) SISCEL (Laboratory Test Control System) designed to monitor laboratory tests, such as CD4 counts and viral load tests for HIV/AIDS patients followed in the public health sector since 2002; (iii) SICLOM (System for Logistic Control of Drugs) developed to control the logistic for the AIDS treatment deliveries; it shares the patients list with SISCEL since 2002. These three databases have been previously combined in a single database with both HIV and AIDS cases using a process called record linkage, which was adopted by the Surveillance Unit of the Brazilian National AIDS Program \citep{Fonseca2010}. This linkage strategy has been increasingly used in AIDS surveillance and research \citep{Deapen2007} to verify under-reporting of cases and eliminate the duplicated ones. In Brazil, that procedure has improved the quality of HIV/AIDS data information \citep{Fonseca2010}. Notice that 2002--2006 can be considered as the first period with substantial information on both HIV/AIDS survival and CD4 exams, where 88 laboratories located in all 27 Brazilian states were using SISCEL, covering 90\% of all CD4 and viral load exams carried out by the public health sector. Cases diagnosed before 2002 were excluded because personal identifiers were not available in the mortality database for the entire country before that date \citep{Fonseca2010}. For institutional reasons, we had access only to a simple random sample of the combined database.

The longitudinal and survival data were collected in a network of 88 laboratories located in every 27 states of Brazil during the years 2002--2006. CD4 counts (a measure of immunologic and virologic status) and survival time after diagnosis were the responses collected in a random sample of $N=500$ individuals, under HAART therapy, from the original data base. While the explanatory variables included were: \verb|age| ($[15,50[$, codded $0$; $\geq 50$, codded $1$), \verb|gender| (Female, codded $0$; Male, codded $1$), \verb|PrevOI| (previous opportunistic infection at study entry, codded $1$; no previous infection, codded $0$), date of HIV/AIDS diagnosis and date of death (available if death happened before 31st December 2006 and censored otherwise). The survival time after diagnosis is calculated as the time period, in years, between date of diagnosis and date of death. As referred in \cite{Souza2007} the variable that accounts for age was chosen basis on the Ministry of Health recommendations, as the over-50 age group showed a higher proportion of delayed initiation of therapy when compared to the population group aged 15-49 years.

There were 34 deaths in this dataset;  440 (88\%) of the patients were between 15 and 49 years old; 298 (59.6\%) were males of whom 23 died. 302 (60.4\%) individuals had no previous infection; 6.4\% lived in the Central-West, 9.6\% in the North-east, 4.6\% in the North, 64.8\% in the South-east region and 14.6\% in the South. The initial median CD4 count was 269 cells/mm$^3$ (men - 250 cells/mm$^3$; women - 295 cells/mm$^3$) and patients made on average 5.51 CD4 exams resulting in a total of 2757 observations. The mean time of a censored patient in the study was 930.2 days (approximately 2.56 years) and for those which had the event that mean time was 862.4 days (approximately 2.36 years).

%-------------------------------------------------------------
\subsection{Fitted models}
\label{sec:fittedmodels}

Our main concern in this work is to use joint modelling for making appropriate inferences about the influence of the longitudinal data characteristics on the survival time. Table \ref{tab:WAIC} shows the form of the 33 joint models fitted to the data and the respective WAIC values for comparison. A square root transformation of the CD4, $\sqrt{\text{CD4}}$, was used in the analysis in order to deal with the high degree of skewness towards the higher counts. Its adjusted longitudinal mean response, $m_i(t_{ij})$, has always the following representation
\begin{equation}
m_{i}(t_{ij}) = \beta_{10} + \beta_{11}t_{ij} + \beta_{12}\verb|sex|_i + \beta_{13}\verb|age|_i + \beta_{14}\verb|PrevOI|_i + b_{1i,1} + b_{1i,2}t_{ij},
\label{eq:longmeanresponseapplied}
\end{equation}
where $\bm{\beta}_1^\top=(\beta_{10},\ldots,\beta_{14})$, $\bm{b}_{1i}^\top=(b_{1i,1},b_{1i,2})$, $\mathbf{x}_{1i}(t_{ij})=(1,t_{ij},\verb|sex|_i,\verb|age|_i,\verb|PrevOI|_i)$ and $\mathbf{w}_{1i}(t_{ij})=(1,t_{ij})$. 

The dispersion model in \eqref{eq:dispersionmodel} will assume several formulations. Supposing that a set of baseline covariates are related to the variability and defining $\bm{\beta}_2^\top=(\beta_{21},\beta_{22},\beta_{23})$, $\bm{b}_{2i}^\top=b_{2i}$, $\mathbf{x}_{2i}(t_{ij})=(\verb|sex|_i,\verb|age|_i,\verb|PrevOI|_i)$ and $\mathbf{w}_{2i}(t_{ij})=1$, we can use the form below to study its effects in the CD4 stability,  
\begin{equation}
\sigma_i^2 = \sigma_0^2\exp\{\beta_{21}\verb|sex|_i + \beta_{22}\verb|age|_i + \beta_{23}\verb|PrevOI|_i + b_{2i}\},
\label{eq:dispersionmodelapplied}
\end{equation}
where $\sigma_0^2$ is the population's common variance and $b_{2i}$
captures different kinds of heterogeneity between individuals, in the same spirit as the frailties in a hazard model. If ones assume that these random effects, $b_{2i}$, are enough to explain the heteroscedasticity, we can define
\begin{equation}
\sigma_i^2 = \sigma_0^2\exp\{ b_{2i} \}.
\label{eq:dispersionmodelappliedonlyb2i}
\end{equation}
More simple forms can be tested, namely considering an individual error variance, $\sigma_i^2$, to which we assign an exchangeable prior distribution, i.e.  
\begin{equation}
\sigma_i^2 | \bm{\theta}_{\sigma^2} \sim \pi(\bm{\theta}_{\sigma^2}),
\label{eq:dispersionmodelexchangeable}
\end{equation}
which means that individual variance, $\sigma_i^2$'s, are i.i.d. random variables conditional on the unknown hyperparameters $\bm{\theta}_{\sigma^2}$. A common error variance can also be assumed, with prior distribution 
\begin{equation}
 \sigma_i^2=\sigma_0^2 \sim \pi(\bm{\theta}_{\sigma^2}), \quad i=1\ldots,N.
\label{eq:commondispersionmodelapplied}
\end{equation}

In the survival part we will always consider the hazard model defined in \eqref{eq:hazard} taking
\begin{equation}
\bm{\beta}_3^\top\mathbf{x}_{3i} = \beta_{31}\verb|sex|_i + \beta_{32}\verb|age|_i + \beta_{33}\verb|PrevOI|_i,
\label{eq:x3applied}
\end{equation}
conjugated with three forms for the function $\mathcal{C}_i(.)$. First a linear one with three individual-specific random effects
\begin{equation}
\mathcal{C}_i(.) = g_1(t)b_{1i,1} + g_2(t)b_{1i,2} + g_3(t)b_{2i},
\label{eq:Ciappliedwith3bi}
\end{equation}
which is the traditional shared random effects approach (\cite{Henderson2000}) extended with $b_{2i}$, which carries information about the individual-specific variability. A form accounting for the explicit effect of the individual longitudinal standard deviation
\begin{equation}
\mathcal{C}_i(.) = g_1(t)b_{1i,1} + g_2(t)b_{1i,2} + g_3(t)\sigma_i,
\label{eq:Ciappliedwithstd}
\end{equation}
which is an approach inspired in \citet{Gao2011} and \citet{McLain2012}. Finally a structure without any information about the residual variability
\begin{equation}
\mathcal{C}_i(.) = g_1(t)b_{1i,1} + g_2(t)b_{1i,2}.
\label{eq:Ciappliedwithoutstd}
\end{equation}
Equations \eqref{eq:Ciappliedwith3bi} and \eqref{eq:Ciappliedwithstd} states that there is some information about the survival process coming from $\sigma_i^2$, i.e. the residual variance may be an indicator of an increase or decrease in the hazard of death. For the time-varying coefficients, $g_1(t)$, $g_2(t)$ and $g_3(t)$ we will use 21 knots (i.e. $s=20$ and $Q=23$) which corresponds to a knot every 3 months during the 5 years. The same is true for $g_0(t)$ when used in \eqref{eq:hazardspline}. When a piecewise constant baseline-hazard is used, and in order to have a term of comparability with $g_0(t)$, we used 20 subintervals, $K=21$, with equal lengths corresponding also to a length of 3 months each during the 5 years.

%-------------------------------------------------------------
\subsection{Prior specification}
\label{sec:priorspecifications}

Under the Bayesian framework we need to inform about the prior specifications for all the unknown parameters. We assume that all the elements of the vectors $\bm{\beta}_1$, $\bm{\beta}_2$ and $\bm{\beta}_3$ are independent of each other and Gaussian distributed, $\mathcal{N}(0,100)$. A Wishart prior distribution, $\mathcal{W}ish(R,\xi)$, was set as the prior for the inverse of the variance-covariance matrix, $\Sigma^{-1}$, of the random effects, $\bm{b}_i$, where $R=\text{diag}(100)$ and $\xi=500/20$ to avoid confounding between the fixed and random effects (\cite{Carlin2001}, p.277). Priors for $\bm{\gamma}_l$, $l=0,1,2,3$, are defined in \eqref{eq:priorrandomwalk}. Generally we will assign a Gamma prior to the inverse of the smoothing parameter, $1/\tau_l^2 \sim \mathcal{G}(0.001,0.001)$, $l=0,1,2,3$.

When the hazard model is defined according to \eqref{eq:hazardpiecewise} we should use some prior specification for the elements of $\bm{\lambda}$, which are unknown. The prior here considered will allow to center the baseline hazard on an Exponential distribution, which has constant hazard, by defining independent Gamma distributions, $\mathcal{G}(\epsilon_{\lambda_k},\epsilon_{\lambda_k})$, on the $\lambda_k$'s, implying that $\mathds{E}[\lambda_k]=1$, i.e. our initial guess is that each piece of the baseline hazard function is constant and equal to 1. We will fix $\epsilon_{\lambda_k}=0.001$, but an hyperprior distribution could be placed on it. Although we should be careful because we can easily end with an over-fitted model. Thus we assume $\lambda_k \sim \mathcal{G}(0.001,0.001)$, $k=1,\ldots,K$. Using small values for $\epsilon_{\lambda_k}$ means larger prior variance for $\lambda_k$ allowing the data to drive departures from the Exponential hazard.

Concerning the dispersion model in \eqref{eq:dispersionmodelapplied} if the linear predictor $\eta_i =\{\beta_{21}\verb|sex|_i + \beta_{22}\verb|age|_i + \beta_{23}\verb|PrevOI|_i + b_{2i}\}=0$, then $\sigma^2_i=\sigma^2_0$ and all the individuals have the same residual variance. Instead if $\eta_i \neq 0$ then we are assuming the existence of some differences in the residual variance behaviour across individuals. Our choice to define the prior for the variance parameter, $\sigma_0^2$, in the scenarios \eqref{eq:dispersionmodelapplied}, \eqref{eq:dispersionmodelappliedonlyb2i} and \eqref{eq:commondispersionmodelapplied} is an approximation to the Jeffreys prior, $\log(\sigma_0) \sim \mathcal{U}(-A,A)$ with $A$ large (100 in this case). The same prior distribution is considered for the individual error variance, $\sigma_i^2$, in \eqref{eq:dispersionmodelexchangeable}. Another two different priors were considered, namely the traditional, $1/\sigma_0^2\sim \mathcal{G}(\epsilon,\epsilon)$ and a half-Cauchy distribution, $\sigma_0|\varpi \sim \text{h-}\mathcal{C}(\varpi)$, $\varpi \sim \mathcal{U}(0,100)$. This later approximation is a natural option when some variances may be outlying \citep{Gelman2006}. It also facilitates the specification of prior beliefs because the only parameter represents the population median standard deviation (\cite{Lunn2013}, p.237). To sample from a half-Cauchy distribution in WinBUGS we have to resort in some distributional calculations, because it's not available in a standard form (for more details \textit{vide} \citet{Lunn2013}, p.225). Although in the results section we only present findings considering that $\log(\sigma_0) \sim \mathcal{U}(-A,A)$, because the conclusions based on the WAIC values and the posterior estimates for the fitted models (Tables \ref{tab:WAIC} and \ref{tab:postestimates}) showed no evidence of being affected by that prior choice.

%-------------------------------------------------------------
\subsection{Results}
\label{sec:results}

Assuming that all the elements of $\bm{\theta}$ are independent of each other, model fitting was performed through MCMC simulation within WinBUGS \citep{Lunn2000} by specifying the full likelihood function and the prior distributions for the unknown parameters. The obtained estimates were based on chains of 500,000 iterations following 250,000 iterations of ``burn-in''. In order to eliminate autocorrelation among samples within the chains, we selected every $25$th iteration of the chains. A study of the trace and density plots of the posterior distributions indicated no convergence problems concerning these samples.

To generate the likelihood when using a piecewise constant function to approximate the baseline hazard as in \eqref{eq:hSpiecewise} we use a ``trick'' that relies on Poisson modelling of expanded data, which means that we are going from an infinite dimensional and unspecified, $h_0(t)$, to a $K$ dimensional parameter vector, $\bm{\lambda}\!=\!(\lambda_1,\ldots,\lambda_K)$. We will not discuss the details here because going into these issues would distract from our main purpose. There exists a broad list of references that can be checked to deepen this subject. The reader should for example consult  \citet[p. 347]{Christensen2011} for the theoretical questions and for the WinBUGS implementation \citet[p. 289]{Lunn2013}. Details about the prior specifications for all the parameters in $\bm{\theta}$ will be given in section \ref{sec:priorspecifications}.

Based on the smallest WAIC value (Table \ref{tab:WAIC}), a P-Spline approach to the baseline hazard seems to be the best strategy. The chosen model $\blacktriangle$ ($\sigma_i^2=$\eqref{eq:dispersionmodelexchangeable}; $\psi_i=$\eqref{eq:x3applied}+\eqref{eq:Ciappliedwithstd}) assumes a heterogeneous within subject variability considering an exchangeable prior distribution  for the $\sigma_i$'s. Additionally, the patient-specific random intercept, slope and standard deviation are considered as covariates for the hazard model. The inclusion of the subject-specific variability as a covariate to explain the survival outcome ($\psi_i=$\eqref{eq:x3applied}+\eqref{eq:Ciappliedwithstd}) improves the results (lower WAIC) compared to the models without this feature ($\psi_i=$\eqref{eq:x3applied}+\eqref{eq:Ciappliedwith3bi} or $\psi_i=$\eqref{eq:x3applied}+\eqref{eq:Ciappliedwithoutstd}). This is scientifically appealing, because the CD4's stability might contain extra information about health-related risks, complementing that provided by the mean-level trajectory. Although, for this particular case, our initial feeling -- a possible effect of some set of covariates related to the individual heterogeneity -- did not show up (models \emptydiamond[0.17]; $\sigma_i^2=$\eqref{eq:dispersionmodelapplied}).

The ``traditional model'' to jointly analyse longitudinal and survival data  \citep{Faucett1996,Henderson2000} considers: (i) a common residual variance as in \eqref{eq:commondispersionmodelapplied}, (ii) a set of two shared random effects ($b_{1i,1}$ and $b_{1i,2}$) believed to drive the relation between the two processes and (iii) the coefficients accounting for the effects of these latent variables on the hazard are time-independent, i.e., $g_1$ and $g_2$ do not depend on $t$, being the function that specifies which components of the longitudinal process are shared with the survival one defined as $\mathcal{C}_i(.) = g_1b_{1i,1} + g_2b_{1i,2}$. In the most part of the literature, the baseline hazard has been tackled considering a Weibull form, but approaches considering splines or a piecewise-constant approximation also have its place \citep{Rizopoulos2011}. Last row of Table \ref{tab:WAIC} has been added so we can compare the performance of these ``traditional'' approaches (models \emptypentagon[0.17]) with our proposed models, namely \emptytriangle[0.17], \emptysquare[0.17] and \emptycircle[0.17]. Considering only the situations where $h_0$ is adjusted using P-Splines, we note that the traditional model \fillpentagon[0.17] is the worst in terms of its WAIC value. 

Table \ref{tab:postestimates} shows the posterior mean estimates together with the 95\% Credible Interval (CI) for a set of parameters for the best model \filltriangle[0.17] compared to its counterparts for the traditional model \fillpentagon[0.17]. One can see that the variables \verb|gender|, \verb|age| and \verb|PrevOI| are all significant in explaining the different longitudinal trajectories and survival times. Older males with opportunistic infections at the study entry have lower $\sqrt{\text{CD4}}$ values (negative estimates). Younger females without opportunistic infections at the study entry have a lower hazard of death (positive estimates). Model \filltriangle[0.17] has been able to capture the effect of the \verb|gender| on the survival outcome as opposed to the model \fillpentagon[0.17]. There is another important thing coming out of these estimates. The unique elements, $\sigma^b_{11}$, $\sigma^b_{12}$ and $\sigma^b_{22}$, of the random-effects covariance matrix, $\Sigma$, for the chosen model are lowered compared to its counterparts for the model \fillpentagon[0.17], which means that we have been able to shrink more the individual random-effects under the presence of a dispersion model for the individual-specific standard deviation than considering a common within-individual variance. This happens because the dispersion model is now explaining some of that additional variability. Some authors, namely \citet{Lyles1999}, have already noted this feature -- random within-subject variances appears to describe a collection of longitudinal biomarker measurements well, as compared to a broad range of alternative models for the correlation structure.

\begin{figure}[!htb]
\centering
\includegraphics[scale=0.5]{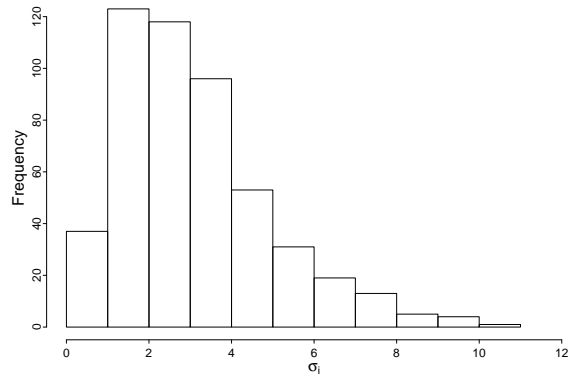}
\caption{Histogram of the posterior means for the $\sigma_i$'s (model \filltriangle[0.17]). There's a considerable heterogeneity among these right skewed  estimates.}
\label{fig:postsigma}
\end{figure}

Figure \ref{fig:postsigma} displays the histogram for the posterior estimates of the within-individual standard deviation, $\sigma_i$, which are considerably heterogeneous and right skewed. This fact is one more reason supporting the assumption of heterogeneous variance which led us to to consider a dispersion model when analysing the CD4 repeated measures. These estimates contrast with the common value, $\sigma_0=2.59$, for the model \fillpentagon[0.17] where a structure for the dispersion is not considered.

\begin{figure}[!htb]
\centering
\includegraphics[scale=1]{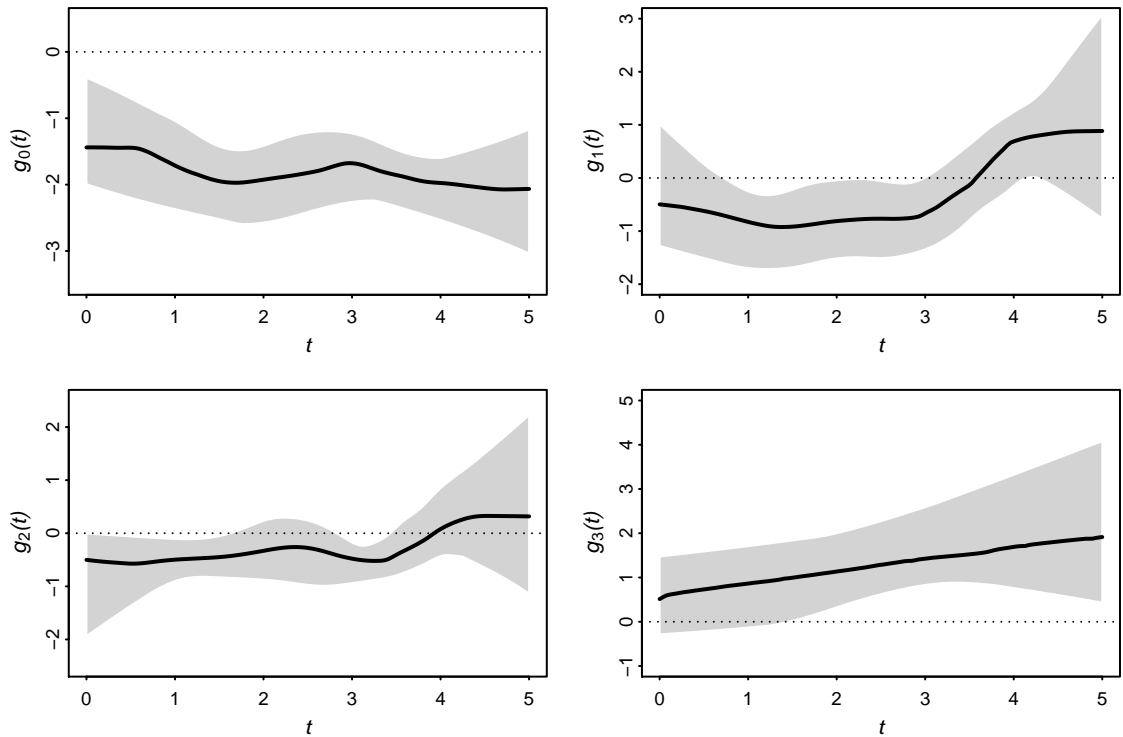}
\caption{Posterior mean estimates, together with the corresponding 95\% Credible Bands (CB), for the selected model \filltriangle[0.17]. The top left panel shows $g_0=\log{(h_0)}$ and the subsequent panels have the time-varying regression coefficients as a function of time in years, $t$. }
\label{fig:gs}
\end{figure}

Figure \ref{fig:gs} top left panel presents $g_0=\log{(h_0)}$. Because its posterior mean (black line) is approximately horizontal and the 95\% Credible Band (CB) lies, approximately, in $(-1,-3)$, we can consider that HIV/AIDS individuals taking medicines correctly have a lower baseline hazard and do not suffer major changes over time. This plot seems to tell us that nowadays people living with HIV, if correctly treated, will have a large life expectancy. Indeed, \citet{May2014} have already noted this in a big study involving 21,388 patients in the UK. Some groups of HIV-positive individuals may expect to live a similar life span to that of the general population, others have reduced life expectancy due to the impact of late diagnosis and late initiation of the therapy. Unfortunately our dataset does not have information about the elapsed time since the infection until the treatment beginning. We only know that older people take longer to start the therapy. The top right panel shows that initially the individual random intercept, $b_{1i,1}$, does not have a significant effect on the hazard but after a short period of time, less than a year, individuals with CD4 counts at the study entry above the mean are at lower risk until, approximately, the begin of the third year when that initial value becomes negligible again. Indeed after a certain period of time it is expected that the initial values become less important. The bottom left panel shows that an individual with a positive time trend for the biomarker has a decreased hazard of death up to 3.5 years after diagnostic, approximately, because the 95\% CB lies below the horizontal dotted line, $g_2(t)=0$. After that period the impact is lowered. Bottom right panel suggests that the higher the within-individual variability the higher the hazard of death and that effect is growing over the time, albeit only after 1.5 years approximately, because the horizontal dotted line, $g_3(t)=0$, is in the 95\% CB before that period. Biologically this is understandable, because if after 1 or 2 years the CD4 counts variability doesn't stabilize, that means that patient's immune system is responding poorly to the treatment. 

Have the possibility to incorporate time-dependent coefficients into a model is often a bonus. Without it we could not have given the interpretations to the coefficients $\bm{g}$ that we gave above. Actually, the posterior estimates for $g_1$ and $g_2$ in model \fillpentagon[0.17] are, respectively, $-0.21$ and $-0.54$, which leads us to gave the following explanation: during the patients follow-up, individuals with higher CD4 values at the study entry and with an upward trend for the CD4 trajectory have a decrease in their risk of death. But, due to the fact that our chosen model \filltriangle[0.17] has time-dependent $g$'s we know now that after a certain period of time both the initial value and the trajectory trend become non-significant.

Unlike the random effects, $\bm{b_i}$, the biomarker variability, $\sigma_i$, has an intuitive meaning for the physicians and respective patients. In a joint model context its interpretation is straightforward, because its contribution to understand the time-to-event can be readily quantified as a hazard ratio (HR). For instance, considering two subjects only differing in their repeated measures variability, everything else being equal for a specific time, the hazard ration is $\text{HR}=\exp\{g_3(t)\}$, which is familiar to the clinical staff.

%=============================================================
\section{Discussion}
\label{sec:discussion}

This paper explores the association between the CD4's trajectory with the time-to-death for patients diagnosed with HIV/AIDS and under HAART therapy, which has important implications in the disease comprehension. The three new aspects presented in this work seem to be improving the traditional framework, videlicet: the dispersion model is capturing some extra-variability in the individual repeated measures implying a reduction in the variability of the individual-specific random effects; the use of the individual level standard deviation as a covariate in the hazard model is scientifically appealing, highlighting that the biomarker's stability might contain extra information about health-related risks, complementing that provided by the mean-level trajectory; the time-dependent coefficients, estimated via P-Splines, that account for the link between the two processes, allow us to understand the influence of the biomarker's values and its variability throughout the time in the survival-related outcome. Obviously, all these results need more studies to support them, but they seem to be encouraging. Beyond AIDS framework, there is an enormous potential to apply the methods developed in this work to other epidemiological contexts.

%==================================================================
\section*{Acknowledgements}

The author wish to thank Maria Goretti Fonseca and Cl\'{a}udia Coeli for the database. This work was partially funded by Egas Moniz, Cooperativa de Ensino Superior.

%================================================================
%\section{References}
\setlength{\bibhang}{1pt}% the hanging indent
\setlength\bibsep{6pt}% the separation b/w basic items
{\footnotesize
\bibliographystyle{chicago}
\bibliography{D:/Research/Bibliography/JM,D:/Research/Bibliography/survival,D:/Research/Bibliography/Bayesian,D:/Research/Bibliography/Longitudinal,D:/Research/Bibliography/DispersionModels,D:/Research/Bibliography/AIDS}
}

%%%%%%%%%%%%%%%%%%%%%%%%%%%%%%%%%%%%%%%%%%%%%%%%%%%%%%%%%%%%%%%%
%%%% TABLES                                                   %%
%%%%%%%%%%%%%%%%%%%%%%%%%%%%%%%%%%%%%%%%%%%%%%%%%%%%%%%%%%%%%%%%

\newpage
%--------- TABLE 1 ( MY FORMAT ) ----------------------------
\begin{table}
\caption{WAIC values for the 33 Bayesian joint dispersion models with flexible links.}
\label{tab:WAIC}
\begin{center}
\footnotesize{
\begin{tabular}[c]{cccclll}
\noalign{\smallskip}\hline\noalign{\smallskip}
%\rowcolor{gray!25}
\multicolumn{2}{c}{\textbf{Longitudinal model}} & & \multicolumn{4}{c}{\textbf{Survival model}} \\\cline{1-2}\cline{4-7}
\noalign{\vspace{0.05cm}}
$m_i$ & $\sigma_i^2$ & & $\varrho_i(t)$ & & $h_0$ & \\\cline{5-7}
    &      &   &    & {Weibull} & {P-Spline} & {Piecewise} \\
\hline
\eqref{eq:longmeanresponseapplied}&\eqref{eq:dispersionmodelapplied} & & \multirow{2}{*}{\eqref{eq:x3applied} + \eqref{eq:Ciappliedwith3bi}} & 14671 & 12573 \emptydiamond[0.17] & 14317 \\
\eqref{eq:longmeanresponseapplied}&\eqref{eq:dispersionmodelappliedonlyb2i} & & & 14700 & 12848 & 14483 \\
\hline
\eqref{eq:longmeanresponseapplied}&\eqref{eq:dispersionmodelapplied} & & \multirow{4}{*}{\eqref{eq:x3applied} + \eqref{eq:Ciappliedwithstd}} & 14307 & 12605 \emptydiamond[0.17] & 13365 \\
\eqref{eq:longmeanresponseapplied}&\eqref{eq:dispersionmodelappliedonlyb2i} & &  & 14452 & 12917 & 13571 \\
\eqref{eq:longmeanresponseapplied}&\eqref{eq:dispersionmodelexchangeable} & & & 13134 \emptytriangle[0.17] & \textbf{12104} \filltriangle[0.17] & 12921 \emptytriangle[0.17] \\
\eqref{eq:longmeanresponseapplied}&\eqref{eq:commondispersionmodelapplied} & & & 13956 \emptysquare[0.17] & 12887 \emptysquare[0.17] & 13533 \emptysquare[0.17] \\
\hline
\eqref{eq:longmeanresponseapplied}&\eqref{eq:dispersionmodelapplied} & &  \multirow{4}{*}{\eqref{eq:x3applied} + \eqref{eq:Ciappliedwithoutstd}} & 14811 & 13334 \emptydiamond[0.17] & 14463 \\
\eqref{eq:longmeanresponseapplied}&\eqref{eq:dispersionmodelappliedonlyb2i} && & 14923 & 13688 & 14599 \\
\eqref{eq:longmeanresponseapplied}&\eqref{eq:dispersionmodelexchangeable} && & 14314 & 13144 & 13968 \\
\eqref{eq:longmeanresponseapplied}&\eqref{eq:commondispersionmodelapplied} & & & 14627 \emptycircle[0.17] & 13553 \emptycircle[0.17]  & 14355 \emptycircle[0.17]  \\
\hline\noalign{\smallskip}
\eqref{eq:longmeanresponseapplied}&\eqref{eq:commondispersionmodelapplied} & & \eqref{eq:x3applied} $\!+g_1b_{1i,1}\! + \!g_2b_{1i,2}$ &   16984 \emptypentagon[0.17]  &  15779 \fillpentagon[0.17] & 16383 \emptypentagon[0.17]  \\
\hline\noalign{\smallskip}
 \end{tabular}}
\end{center}
\end{table}

%O Modelo B8 é o que funciona melhor. ver ficheiro ModelsExplanation.txt

%============================== TABLE 2
\newpage                     %%%%%% MY FORMAT
\begin{table}[!htb]
\caption{Posterior parameters estimates and 95\% Credible Intervals (CI) for the selected model \filltriangle[0.17] and for the traditional joint model \fillpentagon[0.17] with a baseline adjusted with P-Splines.}
\label{tab:postestimates}
\centering
{\footnotesize               %%%%%% MY FORMAT
\begin{tabular}{lrrrr}
\hline
  & \multicolumn{2}{c}{\textbf{Model} \filltriangle[0.17]} &  \multicolumn{2}{c}{\textbf{Model} \fillpentagon[0.17]} \\
%\hline                       %%%%%% MY FORMAT
%\toprule
\textbf{Parameter} & \textbf{Mean} & \textbf{95\% CI} & \textbf{Mean} & \textbf{95\% CI}  \\
\hline                       %%%%%% MY FORMAT
%\midrule
\multicolumn{1}{l}{\underline{\textbf{Longitudinal}:}} \\
Intercept $(\beta_{10})$ & $17.26$  & $(16.59,17.94)$ & $17.3$  & $(16.57,18.02)$  \\
Time      $(\beta_{11})$ & $1.74 $  & $(1.46,2.04)  $ & $1.92$  & $(1.66,2.19)$  \\
Gender    $(\beta_{12})$ & $-0.66$  & $(-1.42,-0.02)$ & $-0.60$ & $(-1.42,0.21)$  \\
Age       $(\beta_{13})$ & $-1.34$  & $(-2.51,-0.41)$ & $-1.59$ & $(-2.58,-0.58)$  \\
PrevOI    $(\beta_{14})$ & $-1.62$  & $(-2.23,-0.99)$ & $-1.80$ & $(-2.77,-0.81)$  \\
$\sigma^b_{11}$          & $20.85$  & $(17.83,24.22)$ & $24.58$ & $(21.32,28.24)$  \\
$\sigma^b_{22}$          & $4.71 $  & $(3.80,5.78)  $ & $6.12$ & $(5.03,7.36)$  \\
$\sigma^b_{12}$          & $-3.07$  & $(-4.46,-1.77)$ & $-4.15$ & $(-5.77,-2.67)$  \\
$\sigma_{0}$           &   $-$    &       $-$       & $2.59$ & $(2.50,2.67)$  \\
\multicolumn{1}{l}{\underline{\textbf{Survival}:}} \\                                                            
Gender   $(\beta_{31})$  & $0.74$  & $(0.49,1.01)$  & $0.49$ & $(-0.30,1.29)$  \\
Age      $(\beta_{32})$  & $0.99$  & $(0.72,1.27)$  & $0.93$ & $(0.04,1.76)$  \\
PrevOI   $(\beta_{33})$  & $1.04$  & $(0.80,1.28)$  & $0.95$ & $(0.19,1.67)$  \\
$g_1$                    &   $-$   &      $-$       & $-0.21$ & $(-0.31,-0.10)$  \\
$g_2$                    &   $-$   &      $-$       & $-0.54$ & $(-0.76,-0.33)$  \\
\hline                       %%%%%% MY FORMAT
%\bottomrule
\end{tabular}
}                            %%%%%% MY FORMAT
\end{table}

%%%%%%%%%%%%%%%%%%%%%%%%%%%%%%%%%%%%%%%%%%%%%%%%%%%%%%%%%%%%%%%%
%%%%  FIGURES                                                 %%
%%%%%%%%%%%%%%%%%%%%%%%%%%%%%%%%%%%%%%%%%%%%%%%%%%%%%%%%%%%%%%%%

%--------- FIGURE 1 ( MY FORMAT ) ----------------------------
\newpage                    

%-------- FIGURE 2 ( MY FORMAT ) -------------------- 

%--------- FIGURE 3 ( MY FORMAT )-------------------

%###################################################################
% ########### SuppMat ####################
\clearpage

%=============================================================
\section{Supplementary Material}
\label{sec:suppmat}

In this supplementary material we provide the detailed version for calculating the WAIC measure (section \ref{sec:WAIC}) and the BUGS code for fitting the selected joint model \filltriangle[0.17] proposed in the paper (section \ref{BUGScode}).

%---------------------------------------------------------------
\subsection{Watanabe-Akaike information criterion (WAIC)}
\label{sec:WAIC}

Following \citet{Gelman2014} we define WAIC as
{\small
\begin{align}
\text{WAIC} = & -2\left[\text{lppd} - p_\text{WAIC}\right] \nonumber \\ 
 = & -2\left[\text{log pointwise predictive density} - \text{ Penalization}\right] \nonumber \\
 = & -2\left[\log	\left\{ \prod_{i=1}^N p(\mathcal{D}_i|\mathcal{D}) \right\}  - \sum_{i=1}^N \mathbb{V}_{\bm{\theta}|\mathcal{D}} \! \left[ \log \{p(\mathcal{D}_i|\bm{\theta})\} \right] \right] \nonumber \\
 = & -2\left[\sum_{i=1}^N \log\left\{ \int p(\mathcal{D}_i|\bm{\theta})\pi(\bm{\theta}|\mathcal{D}) \!\right\} \!d\bm{\theta} \, - \sum_{i=1}^N V_{s=1}^S \left[ \log \{ p(\mathcal{D}_i|\bm{\theta}^s) \} \right]  \right] \nonumber \\
\approx & -2\left[\sum_{i=1}^N \log \left\{ \frac{1}{S} \sum_{s=1}^S p(\mathcal{D}_i | \bm{\theta}^s) \right\} - \sum_{i=1}^N \frac{1}{S-1}
\sum_{s=1}^S \left( \log \{p(\mathcal{D}_i | \bm{\theta}^s)\} - \left\{ \frac{1}{S} \sum_{s=1}^S \log p(\mathcal{D}_i | \bm{\theta}^s) \right\} \right)^{\!2} \right] \nonumber \\
= & -2\left[\sum_{i=1}^N \log \left\{\overline{p(\mathcal{D}_i | \bm{\theta})} \right\} - \sum_{i=1}^N \frac{1}{S-1}
\sum_{s=1}^S \left( \log \{p(\mathcal{D}_i | \bm{\theta}^s)\} - \overline{\log \{p(\mathcal{D}_i | \bm{\theta}) \} } \right)^2 \right] 
\label{eq:WAIC2}
\end{align}}
where $V_{s=1}^S [a_s]$, represents the posterior sample variance and the over line stands for the posterior sample mean. Like the DIC  models with smaller WAIC values  should be preferred.

%==================================================================
\subsection{BUGS code}
\label{BUGScode}

We provide here the BUGS code for fitting the selected joint model \filltriangle[0.17]. The code can be easily generalized to other structures both on the longitudinal and survival parts.

\begingroup
\scriptsize
\begin{verbatim}
#####################################################
#### JOINT DISPERSION MODEL WITH A FLEXIBLE LINK
## Longitudinal model --> linear with subject-specific random effects (eq. 17)
## Dispersion model --> exchangeable prior for the individual variance (eq. 20)
## Hazard model -->  (eq. 4)
## Baseline hazard --> PB-Splines
## Linking structure --> (eq. 24)
## linking parameters g_1(t), g_2(t) and g_3(t) --> time-dependent modeled via PB-Splines
## Penalization through a random walk of order 1.

### Nomenclature
# N - Total number of individuals
# n.exams - Total number of clinical exams per individual
# gender, age and prevoi - covariates
# Y - \sqrt(CD4)
# tauy - precision of the normal distribution
# times - individual observed measurement times 
# beta1 - population parameters for the longitudinal model (\beta_1)
# b - individual random effects
# b0 and tau --> respectively, prior mean and variance for the random-effects
# status - failure indicator, \delta_i
# obs.t <- observed times, T_i
# BS.surv.time - matrix where each line corresponds to each observed time, T_i, evaluated
# for each BSpline basis function
# n.basisfunc - number of BSpline basis functions (Q)
# beta3 - \beta_3
# gamma.0, gamma.1, gamma.2 and gamma.3 --> \gamma coefficients in eq. (5)
# eta_i --> eq. 22 + eq. 24

model{

C <- 10000 # constant to use in the zeros trick

############################
### Longitudinal model
############################
for(i in 1:N){ 
    for(j in 1:n.exams[i]) {
        Y[i,j] ~ dnorm(muy[i,j], tauy[i])
        muy[i,j] <- beta1[1] + beta1[2]*times[i,j] + beta1[3]*gender[i] + beta1[4]*age[i] + beta1[5]*prevoi[i] +
                    b[i, 1] + b[i,2]*times[i,j] 

        # log(L_1) = Log-Likelihood longitudinal part for individual i at time j
        log_like_long[i,j] <- log(sqrt(tauy[i]/6.2831)) - tauy[i]*0.5*pow(Y[i,j]-muy[i,j],2)
    
        # log(L_1*L_2) = full log-likelihood for individual i at time j (eq. 12)
        log_like[i,j] <- log_like_long[i,j] + log_like_surv[i]  
       }

    # Random effects
    b[i,1:2] ~ dmnorm(b0[], tau[,])  
  }

       
################################
### Survival model
################################

## Calculate \varrho_i(t) at T_i (eq.22 + eq. 24)
for(i in 1:N) {
    varrho[i] <- beta3[1]*gender[i] + beta3[2]*age[i] + beta3[3]*prevoi[i] +
                 g1[i]*b[i,1] + g2[i]*b[i,2] + g3[i]*sigmay[i]

    ## Calculate the value of the splines g0(.)=log[h_0], g1(.), g2(.) and g3(.)
    ## at T_i (obs.t[i]) --> g0[i], g1[i], g2[i] and g3[i]
    for(k in 1:n.basisfunc){
        BS.g0[i, k] <- gamma.0[k] * BS.surv.time[i, k]
        BS.g1[i, k] <- gamma.1[k] * BS.surv.time[i, k]
        BS.g2[i, k] <- gamma.2[k] * BS.surv.time[i, k]
        BS.g3[i, k] <- gamma.3[k] * BS.surv.time[i, k]
       }  
    g0[i] <- sum(BS.g0[i, 1:n.basisfunc])        ## log-baseline hazard --> log[h_0] = g_0(t)
    h0[i] <- exp(g0[i])                          ## baseline hazard
    g1[i] <- sum(BS.g1[i, 1:n.basisfunc])
    g2[i] <- sum(BS.g2[i, 1:n.basisfunc])
    g3[i] <- sum(BS.g3[i, 1:n.basisfunc])

    log.hazard[i] <- (g0[i] + eta[i])*varrho[i] 
   }


## Calculate the integral of the hazard function, I=\int_0^{T_i}{ \exp{g_0(u)+\varrho_i(u) du } }
# Using the Gauss-Kronrod quadrature rule
# phi - the 15 Gauss-Kronrod quadrature points
# phi <- c(-0.949107912342758524526189684047851, -0.741531185599394439863864773280788,
#          -0.405845151377397166906606412076961, 0, 0.405845151377397166906606412076961,
#          0.741531185599394439863864773280788,  0.949107912342758524526189684047851,
#          -0.991455371120812639206854697526329, -0.864864423359769072789712788640926,
#          -0.586087235467691130294144838258730, -0.207784955007898467600689403773245,
#          0.207784955007898467600689403773245,  0.586087235467691130294144838258730,
#          0.864864423359769072789712788640926,  0.991455371120812639206854697526329)
# w - the 15 Gauss-Kronrod quadrature weights
# w <- c(0.063092092629978553290700663189204, 0.140653259715525918745189590510238,
#      0.190350578064785409913256402421014, 0.209482141084727828012999174891714,
#      0.190350578064785409913256402421014, 0.140653259715525918745189590510238,
#      0.063092092629978553290700663189204, 0.022935322010529224963732008058970,
#      0.104790010322250183839876322541518, 0.169004726639267902826583426598550,
#      0.204432940075298892414161999234649, 0.204432940075298892414161999234649,
#      0.169004726639267902826583426598550, 0.104790010322250183839876322541518,
#      0.022935322010529224963732008058970)



# loop for the Gauss-Kronrod quadrature rule
for(i in 1:N){  
    # P1=(b-a)/2 = (obs.t[i] - 0)/2 and P2=(obs.t[i] + 0)/2 
    # in this case P1=P2 because the lower bound of the integral is 0=a
    P1[i] <- obs.t[i]/2
    P2[i] <- P1[i]
    # loop to calculate the value of the splines g0, g1, g2 and g3 at the 15 quadrature points, "phi"
    # u is the value (obs.t_i/2)*phi + obs.t_i/2
    # BS.u.vect is a matrix which dim=(15*500)x(n.basisfunc);
    # each line is the transformation of the quadrature points (phi) to the B-Spline Basis 
    # varrho.u is the value of varrho at point u 
    for (j in 1:15) {
         for(k in 1:n.basisfunc){
             BS.g0.u[i, j, k] <- gamma.0[k] * BS.u.vect[i*15-15+j, k]
             BS.g1.u[i, j, k] <- gamma.1[k] * BS.u.vect[i*15-15+j, k]
             BS.g2.u[i, j, k] <- gamma.2[k] * BS.u.vect[i*15-15+j, k]
             BS.g3.u[i, j, k] <- gamma.3[k] * BS.u.vect[i*15-15+j, k]
            }
         g0.u[i, j] <- sum(BS.g0.u[i, j, 1:n.basisfunc])
         g1.u[i, j] <- sum(BS.g1.u[i, j, 1:n.basisfunc])
         g2.u[i, j] <- sum(BS.g2.u[i, j, 1:n.basisfunc])
         g3.u[i, j] <- sum(BS.g3.u[i, j, 1:n.basisfunc])

         varrho.u[i,j] <- beta3[1]*gender[i] + beta3[2]*age[i] + beta3[3]*prevoi[i] +
                          g1.u[i, j]*b[i,1] + g2.u[i, j]*b[i,2] + g3.u[i, j]*sigmay[i]
         
         w.exp.g0.varrho[i, j] <- w[j] * exp(g0.u[i, j] + varrho.u[i, j])   # terms to sum 
         
        }

    log.survival[i] <- -P1[i] * sum(w.exp.g0.varrho[i, 1:15])
    # log-Survival; -\int_0^{T_i}{ e^{g0(u)+varrho_i(u)du} } aproximates the minus integral, i.e.,
    # the (-cumulative hazard), -H, summing over the 15 quadrature points  

    log_like_surv[i] <- log.hazard[i] + log.survival[i]   # log(L_2) = Log-Likelihood survival part

    # To define the likelihood of the survival model we use the zeros trick
    # of WinBUGS, where C is a positive constant
    zeros[i] <- 0
    zeta[i] <- - log_like_surv[i] + C
    zeros[i] ~ dpois(zeta[i])
   }

############################
### Priors

## Penalized prior for the coefficients gamma.0, gamma.1, gamma.2 and gamma.3
## of the functions g0, g1, g2 and g3
# random walk prior of order 1
gamma.0[1] ~ dnorm(0, 0.01)  
gamma.1[1] ~ dnorm(0, 0.01)
gamma.2[1] ~ dnorm(0, 0.01) 
gamma.3[1] ~ dnorm(0, 0.01) 
for(k in 2:n.basisfunc) {
    gamma.0[k] ~ dnorm(gamma.0[k-1], tau.g0)
    gamma.1[k] ~ dnorm(gamma.1[k-1], tau.g1)
    gamma.2[k] ~ dnorm(gamma.2[k-1], tau.g2)
    gamma.3[k] ~ dnorm(gamma.3[k-1], tau.g3)
   }

## prior for the individual variance 
for(i in 1:N){
   log.sigmay[i] ~ dunif(-100,100) 
   sigmay2[i] <- pow(exp(log.sigmay[i]), 2)   
   tauy[i] <- 1/sigmay2[i]
   sigmay[i] <- sqrt(sigmay2[i])            # std. deviation of the longitudinal measures for the i individual
}

## other priors
tau.g0 ~ dgamma(0.01, 0.01)
tau.g1 ~ dgamma(0.01, 0.01)
tau.g2 ~ dgamma(0.01, 0.01)
tau.g3 ~ dgamma(0.01, 0.01)
for(k in 1:5) { beta1[k] ~ dnorm(0, 0.01) }
for(k in 1:3) { beta3[k] ~ dnorm(0, 0.01) }
tau[1:2,1:2] ~ dwish(V[,], 25) 
}
# end of the model
\end{verbatim}
\endgroup

%===============================================================
\end{document}